\documentclass[preprint,review,1p,times]{elsarticle}
\usepackage{gensymb}
\usepackage{xcolor}
\usepackage{comment}
\usepackage{xurl}
\usepackage{float}
\usepackage{placeins}  

\begin{document}
\begin{frontmatter}

\title{Development of a Silicon-Based Ultra-Fast X-Ray Beam Size Monitor for SuperKEKB}

\author[uhm]{Matthew Andrew}
\author[utokyo,kek]{Riku Nomaru}
\author[kek,sokendai]{Gaku Mitsuka}
\author[uhm]{Keisuke Yoshihara\corref{cor1}}
\ead{kyoshiha@hawaii.edu}
\author[uhm]{Cody Driver}
\author[slac]{Christopher J. Kenny}
\author[nagoya]{Bela Urbschat}
\author[uhm]{Gary Varner}
\author[kek,sokendai]{John W. Flanagan}

\cortext[cor1]{Corresponding author.}

\address[uhm]{University of Hawaii, Honolulu, Hawaii 96822, USA}
\address[kek]{High Energy Accelerator Research Organization (KEK), Tsukuba 305-0801, Japan}
\address[sokendai]{The Graduate University for Advanced Studies (SOKENDAI), Hayama 240-0193, Japan}
\address[utokyo]{Department of Physics, University of Tokyo, Tokyo 113-0033, Japan}
\address[slac]{SLAC National Accelerator Laboratory, 2575 Sand Hill Road, M/S 96, Menlo Park, CA 94025, USA}
\address[nagoya]{Graduate School of Science, Nagoya University, Nagoya 464-8602, Japan}

\begin{abstract}
We present the development of a silicon-based ultra-fast X-ray beam size monitor (SiXRM) for SuperKEKB. The system enables, for the first time at SuperKEKB, bunch-by-bunch measurements of the vertical beam size using synchrotron radiation. The detector combines a silicon strip sensor board, amplifier boards, and fast waveform readout electronics. Beam measurements demonstrate clear reconstruction of the bunch structure and X-ray images for individual bunches. The measured beam sizes show good agreement with those obtained from the existing CMOS-based XRM system. The measurement precision is estimated to be better than 6.4~$\mu$m. This system provides a powerful tool for studying beam dynamics and optimizing luminosity in high-luminosity colliders.
\end{abstract}


\end{frontmatter}

\section{Introduction} 

SuperKEKB~\cite{Ohnishi2013} is a high-luminosity asymmetric-energy electron-positron collider located at KEK in Tsukuba, Japan. It operates at center-of-mass energies corresponding to the $\Upsilon$(4S) resonance and provides electron--positron collisions for the Belle~II experiment~\cite{Abe2010,Adachi2018,Kou2019}. SuperKEKB adopts the nanobeam scheme, requiring extremely small vertical beam sizes ($\beta_y^{*}=0.3$~mm) at the interaction point (IP) to reach the target luminosity of $6 \times 10^{35}$~cm$^{-2}$~s$^{-1}$. Since the start of Phase~3 operation in 2019, SuperKEKB has steadily increased the stored beam currents and delivered progressively higher luminosity. On May 30, 2026, the machine achieved a peak luminosity of $5.281 \times 10^{34}$~cm$^{-2}$~s$^{-1}$, setting a new world record for $e^{+}e^{-}$ colliders~\cite{SuperKEKBOperation2026}.

Precise measurement of the vertical beam size in both the electron (HER) and positron (LER) rings is essential for luminosity optimization, beam tuning, and machine stability studies. Because the beam conditions can vary from bunch to bunch within a stored bunch train, a diagnostic that resolves individual bunches provides information that is lost when the signal is averaged over the train. Such measurements are particularly important for studying bunch-dependent beam blow-up driven by collective and single-bunch beam dynamics.

Conventional synchrotron radiation monitors based on visible light are widely used for beam diagnostics; however, their spatial resolution is fundamentally limited by diffraction, preventing measurements of vertical beam sizes below $\sim$10~$\mu$m. X-ray beam size monitors offer improved spatial resolution and have been used at SuperKEKB with CMOS cameras~\cite{Mulyani:2019gsy}. In such systems, the camera image is formed by accumulating X-ray photons over the camera integration time. The reconstructed beam size therefore represents an average over the bunches contributing to the image. Consequently, bunch-to-bunch variations of the vertical beam size along the train cannot be resolved. At SuperKEKB, where the minimum bunch spacing is as short as 4~ns, resolving individual bunches requires a detector system with sufficiently fast temporal response and waveform readout capability.

To overcome this limitation, a silicon-based ultra-fast X-ray beam size monitor (SiXRM) has been developed for SuperKEKB. Silicon sensors provide fast response and fine spatial granularity, making them well suited for bunch-resolved X-ray detection. The SiXRM employs a silicon strip sensor and dedicated fast waveform readout electronics to reconstruct the X-ray image for each bunch and extract the corresponding vertical beam size. The system has been installed in the HER for initial operation.

This capability provides direct insight into beam-size evolution along the bunch train and supports machine studies aimed at maximizing luminosity and mitigating beam instabilities. In this paper, we describe the design and operation of the SiXRM system and present the first beam-size measurements obtained during SuperKEKB operation.

\section{X-Ray Beam Size Measurement} 
\label{sec:xray_beam_size_measurement}

The principles of X-ray beam size measurement are described in detail in Ref.~\cite{Mulyani:2019gsy}. In this paper, X-ray radiation monitor (XRM) refers to the coded-aperture X-ray beam size monitor used at SuperKEKB. The SiXRM shares the same optical system as the existing CMOS-based XRM, with the SiXRM detector installed downstream of the CMOS camera. This configuration allows the reuse of the coded aperture response templates and enables direct cross-checks of the measured beam size between the two systems.

\begin{figure}[h]
  \centering
  \includegraphics[width=0.85\linewidth]{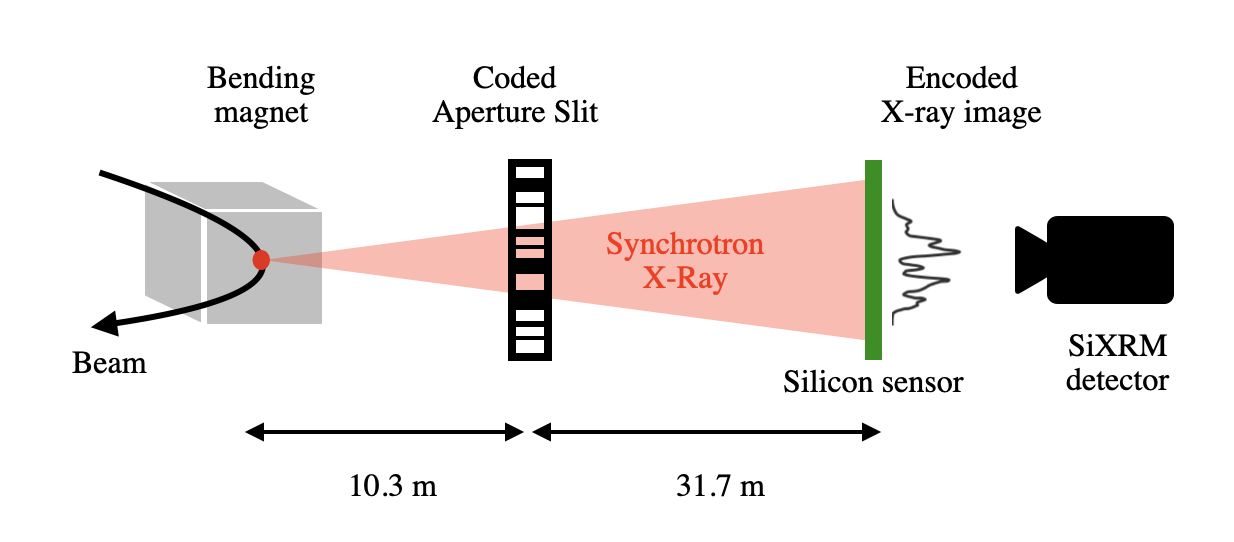}
  \caption{Conceptual layout of the SiXRM beam size measurement. Synchrotron X-rays emitted from a bending magnet pass through a coded aperture located 10.3~m downstream of the source. The detector box is located 31.7~m downstream of the coded aperture, where the encoded X-ray image is detected by the silicon sensor and read out by the SiXRM detector system.}
  \label{fig:sixrm_measurement_concept}
\end{figure}

\subsection{Synchrotron radiation source}

The SiXRM utilizes synchrotron radiation emitted from bending magnets. As charged particles traverse a curved trajectory in a magnetic field, they emit broadband radiation extending into the X-ray regime.

The spectrum is continuous and characterized by the critical photon energy, which depends on the beam energy and bending radius. At SuperKEKB, the critical energies are approximately 4.4~keV (LER) and 7.2~keV (HER), resulting in X-ray spectra spanning from sub-keV to several tens of keV. The HER produces a harder spectrum with a larger fraction of high-energy photons.

As the X-rays propagate along the beamline, they undergo energy-dependent attenuation in windows, air, and optical components. The spectrum incident on the detector is therefore softer or harder depending on the beamline transmission. The transmitted X-rays are directly detected by a silicon strip sensor, where absorbed photons generate charge carriers. The number of detected photons and the deposited charge per photon determine the signal amplitude and statistical precision of the beam-size measurement.

\subsection{X-ray imaging and beam size reconstruction}

X-rays from the synchrotron radiation source pass through a coded aperture and are imaged onto the detector. A coded aperture consists of a patterned array of multiple openings, allowing higher photon throughput compared to a single-aperture (e.g., pinhole) configuration while encoding spatial information in the detected intensity distribution. The image formed by a point-like source is not a geometrical projection of the aperture alone; it is broadened by diffraction and defines the point spread function of the optical system.

For a point-like source, the diffraction pattern formed on the detector after the coded aperture defines the point spread function (PSF) of the system. The PSF depends on the X-ray spectrum, beamline geometry, and aperture pattern. For a beam with a finite transverse size, the measured image is given by the convolution of the PSF with the beam profile.

\begin{figure}[H]
  \centering
  \includegraphics[width=0.75\linewidth]{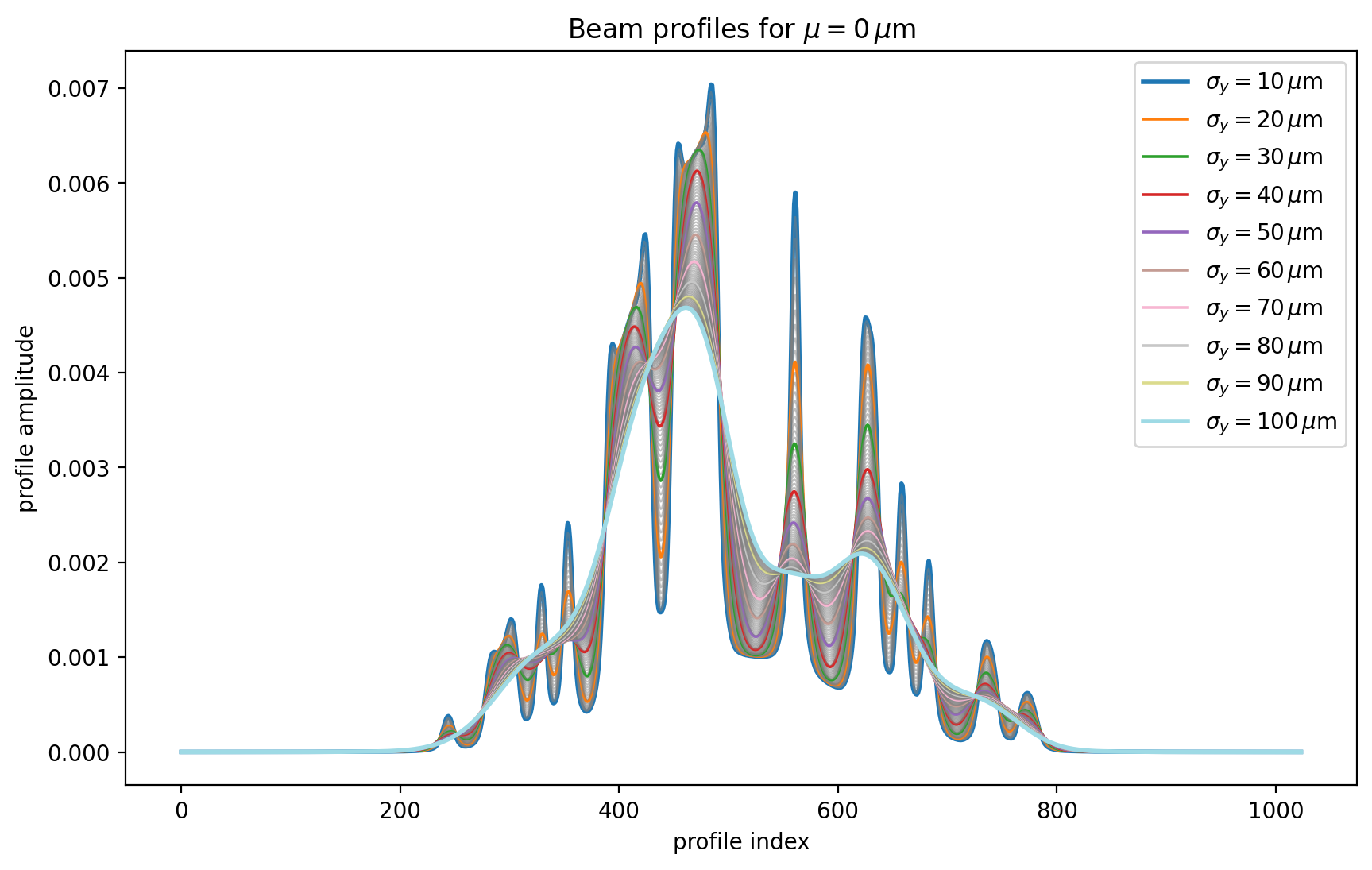}
  \caption{Example of simulated X-ray image templates for a fixed vertical beam position of $\mu=0~\mu\mathrm{m}$ and different assumed vertical beam sizes $\sigma_y$. The horizontal axis is the profile index, corresponding to the sampled position in the one-dimensional detector image, and the vertical axis is the normalized profile amplitude. As $\sigma_y$ increases, the coded-aperture image becomes broader and the fine diffraction structures are smoothed.}
  \label{fig:template_profiles_mu0}
\end{figure}

The vertical beam size is determined by fitting the measured image to a set of pre-calculated templates. These templates are generated using simulations based on the PSF and detector response for different assumed beam sizes. The source beam profile is assumed to be Gaussian, and templates are prepared by convolving the PSF with Gaussian beam profiles of different vertical sizes $\sigma_y$. The PSF and detector-response model include the effects of the X-ray spectrum, beamline transmission, and detector response. An example set of templates is shown in Fig.~\ref{fig:template_profiles_mu0}. The fit is performed using a least-squares procedure, where the beam size is extracted together with nuisance parameters describing image position offsets and amplitude scaling.

The use of a coded aperture improves photon statistics, while the template-based approach enables robust beam size determination even for structured detector images.

\FloatBarrier

\section{Detector System Overview}

\begin{figure}[h]
  \centering
  \includegraphics[width=0.8\linewidth]{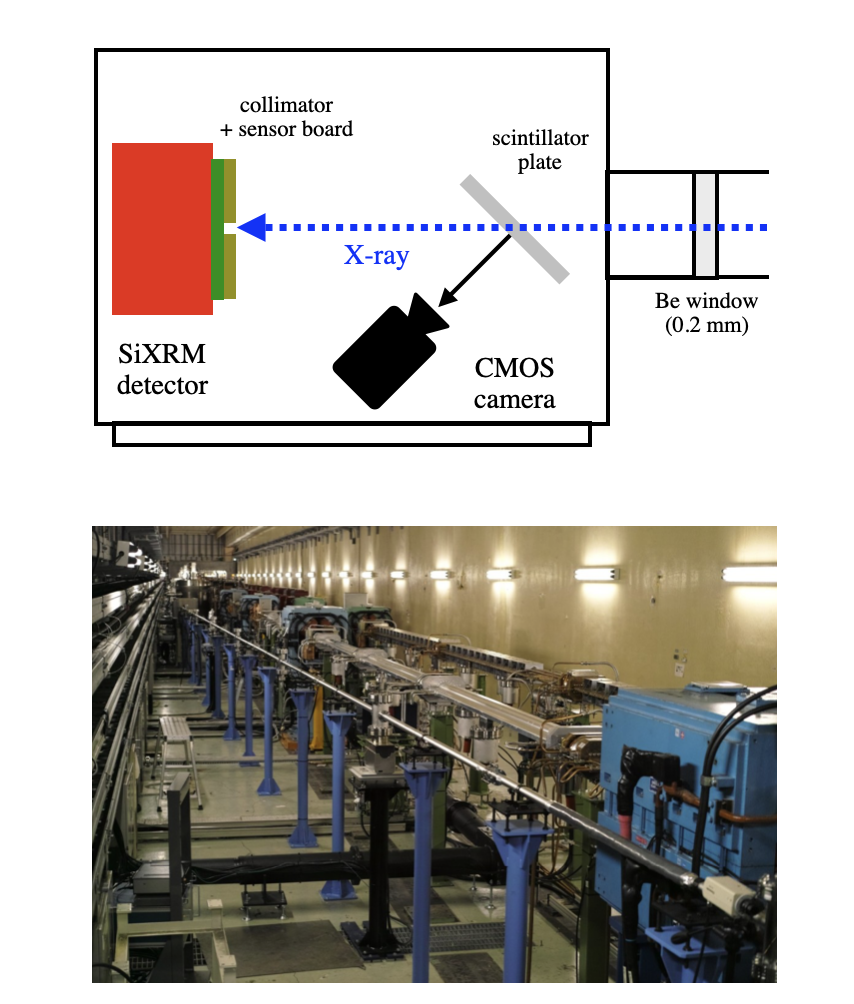}
  \caption{Overview of the X-ray detector system. Top: schematic of the detector box showing the CMOS-based XRM and the downstream SiXRM. Bottom: photograph of the SuperKEKB tunnel with the X-ray transport vacuum pipe.}
  \label{fig:detector_box}
\end{figure}

The SiXRM detector system is installed in the X-ray beamline at the D04 section of the SuperKEKB tunnel, as illustrated in Fig.~\ref{fig:detector_box}. After passing through the coded-aperture optics described in Sec.~\ref{sec:xray_beam_size_measurement}, the X-rays are transported through a dedicated vacuum pipe and enter the detector box through a Be window.

The detector box houses both the conventional CMOS-based XRM and the SiXRM detector, allowing them to share the same optical path. The upstream CMOS system uses a scintillator plate and optical imaging, while the downstream SiXRM directly detects transmitted X-rays using a silicon strip sensor with a copper collimator. This configuration enables direct comparison and cross-calibration between the two measurement systems. 

\begin{figure}[h]
  \centering
  \includegraphics[width=0.8\linewidth]{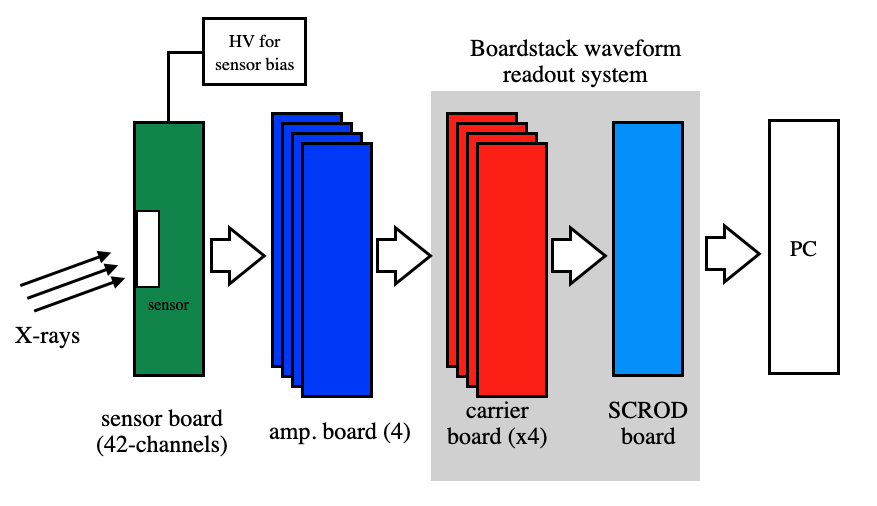}
  \caption{Block diagram of the SiXRM detector system. Signals from the silicon sensor are amplified, digitized, and processed using a boardstack-based readout system, and transferred to a readout PC.}
  \label{fig:readout_overview}
\end{figure}

The SiXRM detector system consists of a silicon strip sensor board, amplifier boards, and a boardstack-based readout system, forming a compact and modular readout chain, as illustrated in Fig.~\ref{fig:readout_overview}. Synchrotron X-rays are detected by a silicon strip sensor mounted on the sensor board (42 channels) and converted into electrical signals, which are subsequently amplified by wideband amplifier boards prior to digitization.

The amplified signals are digitized and processed using a boardstack-based waveform readout system originally developed for the Belle~II Time-of-Propagation (TOP) front-end electronics~\cite{Kotchetkov2019TOP}. In the SiXRM system, this boardstack is reused as a compact multi-channel platform for waveform digitization, control, and data aggregation. The detailed readout architecture is described in Sec.~\ref{sec:waveform_readout_system}.

The firmware is based on that of the original TOP system and has been modified to meet the requirements of the SiXRM application. The processed data are transferred to a readout PC via a high-speed link.

Calibration and commissioning studies were carried out at the testbench in the D04 ground hut using a dedicated laser system; details are described in Sec.~\ref{sec:Gain_Calibration}.

\subsection{Silicon Sensor and Sensor Board} 

The detector employs a silicon strip sensor fabricated at the Stanford Nanofabrication Facility. The sensor features 128 cathode strips with a pitch of 50~$\mu$m. The depletion depth is approximately 75~$\mu$m, and the sensor is typically operated at a bias voltage of -40~V.

\begin{figure}[h]
  \centering
  \includegraphics[width=0.8\linewidth]{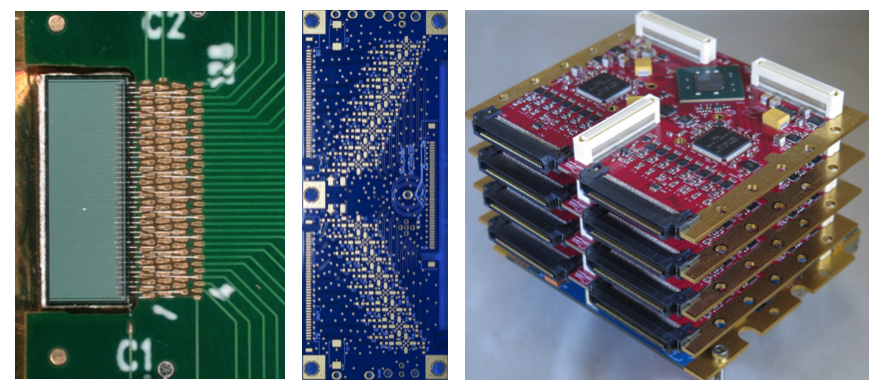}
  \caption{Photographs of the SiXRM readout components. Left: silicon strip sensor mounted on the sensor board with wire-bonded channels. Center: amplifier board for signal conditioning. Right: boardstack-based waveform readout system consisting of multiple carrier boards and a SCROD board.}
  \label{fig:readout_pcb}
\end{figure}

For the present study, a subset of 42 channels out of the 128 available strips is wire-bonded to a custom-designed sensor board, as shown in Fig.~\ref{fig:readout_pcb} (left). The sensor board provides signal routing and bias distribution for the sensor. The wire-bonded region is fixed with epoxy to ensure mechanical stability, and a copper support plate is attached to the backside of the board to provide structural rigidity.

\subsection{Amplifier Board} 

The sensor signals are read out using four amplifier boards, each handling 32 channels out of the total 128 sensor channels. This configuration follows the original segmentation of the sensor and allows scalable readout of the full channel count. A photograph of the amplifier board is shown in Fig.~\ref{fig:readout_pcb} (center).

In the SiXRM system, the signal amplitude from the silicon sensor is relatively small due to the limited charge deposited by X-rays. Each amplifier channel uses a Mini-Circuits MAR-6SM+ surface-mount monolithic wideband amplifier, which provides an approximate gain of 20~dB~\cite{MiniCircuits_MAR6SM}. The amplifier boards are therefore designed to provide sufficient gain with fast response, enabling accurate waveform reconstruction while preserving timing information.

\subsection{Waveform readout system} 
\label{sec:waveform_readout_system}

For the waveform readout system, we reuse the TOP detector boardstack developed for the Belle~II experiment, which provides a proven architecture for high-speed waveform acquisition. A single boardstack consists of four carrier boards and one Standard Control Read-Out Data (SCROD) board, forming a modular readout unit with a total of 128 input channels. The boardstack-based readout system is shown in Fig.~\ref{fig:readout_pcb} (right).

\subsubsection{IRSX waveform digitizer ASICs}

The Ice Ray Sampler X (IRSX) is an 8-channel multi-gigasample-per-second waveform digitizer ASIC, fabricated in a 0.25~$\mu$m CMOS process. It employs a switched capacitor array (SCA) architecture for analog waveform sampling at a rate of 2.7~Gsps, in which sampling and digitization are decoupled: analog waveforms are continuously stored in on-chip capacitors and digitized subsequently. Each channel contains 32,768 analog storage cells, corresponding to a buffer depth of approximately 12~$\mu$s at the nominal sampling rate. Each storage cell consists of a small capacitor and a comparator, enabling sample-and-hold operation and selective digitization of stored signals.

The digitization is based on a Wilkinson analog-to-digital conversion scheme. A common ramp voltage is applied, and the conversion time is encoded using a Gray code counter. When the ramp voltage exceeds the stored sample voltage, a comparator triggers and latches the corresponding counter value. This approach allows compact implementation with a typical resolution of 12 bits over a $\sim$1.5~V dynamic range.

\subsubsection{Carrier boards}

Each carrier board hosts four IRSX ASICs, each providing 8 input channels, resulting in 32 channels per carrier and 128 channels per boardstack. Each carrier board also includes two voltage-sensitive amplifiers to condition the input signals prior to sampling. The boards are equipped with AMD/Xilinx Zynq-7000 series devices for local control, buffering, and configuration.

The carrier-board firmware is largely the same as that used in the TOP application.

\subsubsection{SCROD board}

Digitized data are transmitted from each carrier to the SCROD board via LVDS-based serial links. The SCROD board, also equipped with an AMD/Xilinx Zynq-7000 series device, acts as a data concentrator, aggregating the data streams from the four carriers into a unified output stream. The FPGA on the SCROD serves as the main platform for event processing and digital signal processing in the readout chain.

In the original TOP system, additional processing such as feature extraction is performed in the processing system (PS). In contrast, the XRM application reduces the use of PS-side processing on the SCROD; feature extraction is not performed online, and the SCROD is primarily used for data aggregation and event building.

The aggregated data are then transferred to a backend readout PC via a high-speed optical link for further processing and storage.

\subsubsection{Commissioning of readout chain}

During prototype evaluation, oscillations attributed to cross-talk between carrier boards were observed when the inter-board spacing was 9~mm. Increasing the spacing to 16~mm successfully suppressed this oscillatory behavior, indicating that sufficient physical separation is essential to mitigate cross-talk-induced instabilities.

In the current implementation, the XRM application does not rely on an external L1 trigger. Full waveform information is preserved and transferred to the backend readout PC. Data acquisition (DAQ) is based on a sweep readout scheme, in which stored analog samples are digitized sequentially. The beam size is then reconstructed using offline analysis software.

\subsection{University of Hawaii testbench} 

A dedicated testbench has been developed at the University of Hawaii for electronics verification prior to deployment at KEK. The setup enables standalone testing of the full readout chain, including firmware and DAQ system.

The testbench is based on a sensor emulator board with FPGA-generated signals, providing up to 64 channels and reproducing the 42-channel SiXRM configuration. Signals are sent to the readout boardstack (carrier boards and SCROD) and then transferred to a readout PC, enabling end-to-end system validation. This setup allows controlled multi-channel testing of timing, synchronization, and event building.

\section{Vertical Beam Size Reconstruction} 

Bunch-resolved X-ray images and vertical beam sizes are reconstructed from silicon strip waveforms through three principal steps: channel gain calibration, pulse-height extraction from beam waveforms, and template fitting of the reconstructed X-ray images.

\subsection{Gain Calibration}
\label{sec:Gain_Calibration}
Relative calibration coefficients are determined before beam measurements to correct for channel-to-channel variations in the response of the silicon sensor and readout chain.
A dedicated calibration testbench is installed in the D04 ground hut, outside the accelerator tunnel.
The SiXRM system and a pulsed laser are housed in a light-tight enclosure, with the laser mounted on a motorized vertical stage so that its illumination position can be controlled remotely.
The acquired data are transmitted to an external computer via optical fiber, and the sensor bias voltage is also controlled remotely.

\begin{figure}[h]
  \centering
  \begin{minipage}[b]{0.24\linewidth}
    \centering
    \includegraphics[height=0.27\textheight]{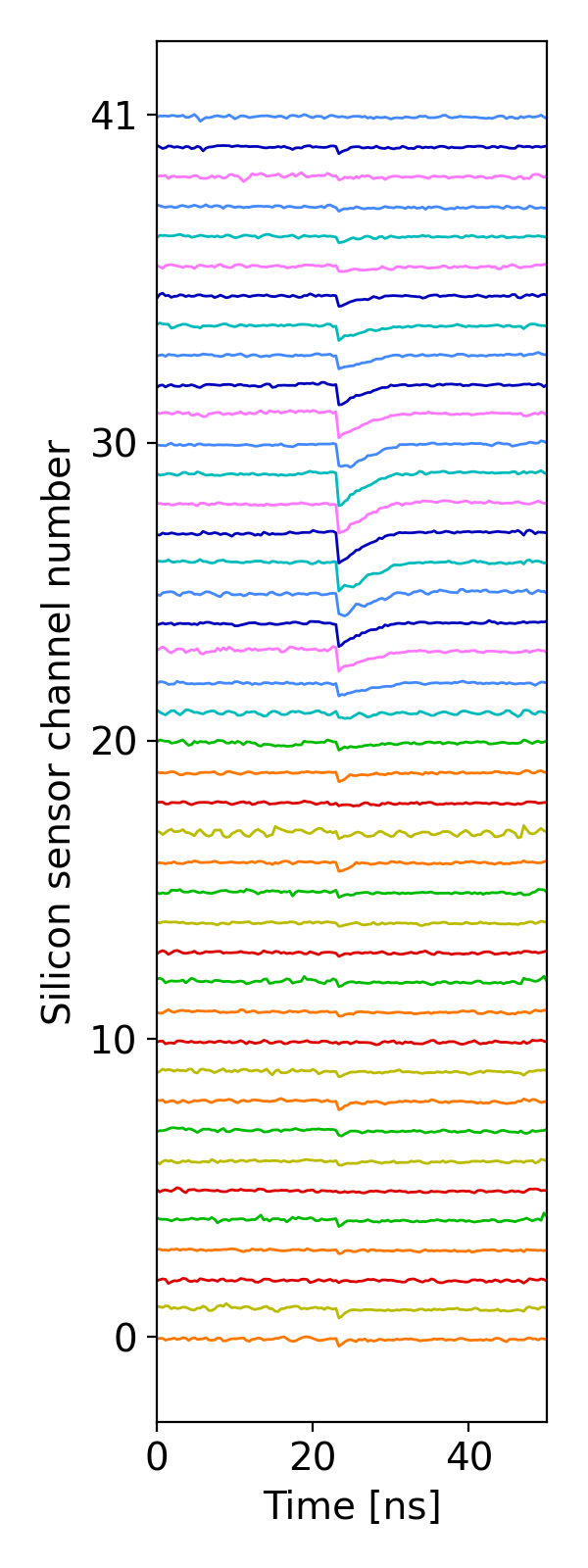}\\
    (a)
  \end{minipage}
  \hfill
  \begin{minipage}[b]{0.66\linewidth}
    \centering
    \includegraphics[height=0.27\textheight]{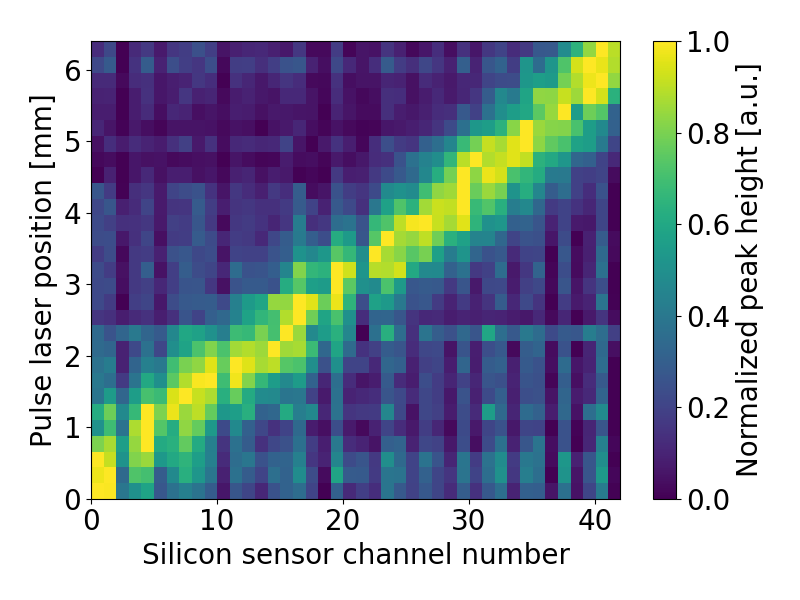}\\
    (b)
  \end{minipage}
  \caption{Laser-based gain-calibration measurements: (a) waveforms recorded from sensor channels 0--41 in response to a single laser pulse; (b) normalized pulse height as a function of sensor channel and vertical laser position.}
  \label{fig:gain_calibration}
\end{figure}

The calibration uses a Thorlabs NPL98B pulsed laser~\cite{Thorlabs_NPL98B}, which emits light at a wavelength of $980 \pm 10\,\mathrm{nm}$ and provides five selectable pulse widths.
The shortest available pulse width, approximately 6~ns, is used for these measurements.
Figure~\ref{fig:gain_calibration}(a) shows a typical SiXRM response to a single laser pulse, for which signals are observed primarily in channels 20--40.
To characterize the response of all channels, the laser is scanned vertically across the sensor, and waveform data are acquired at each position.
The resulting scan, shown in Fig.~\ref{fig:gain_calibration}(b), exhibits a linear shift of the responding channels with laser position, confirming the spatial correspondence between the illumination position and sensor channel.
The pulse heights are normalized to the range 0--1 at each laser position for visualization.
Assuming stable laser intensity during the scan, the measured channel responses are used to determine the relative calibration coefficients.
Because the response also depends on the particular combination of sensor channel, preamplifier, and DAQ board, the coefficients are re-evaluated whenever the readout configuration is changed.

\subsection{Beam Measurement}
\begin{figure}[H]
  \centering
  \begin{minipage}[b]{0.72\linewidth}
    \centering
    \includegraphics[width=\linewidth]{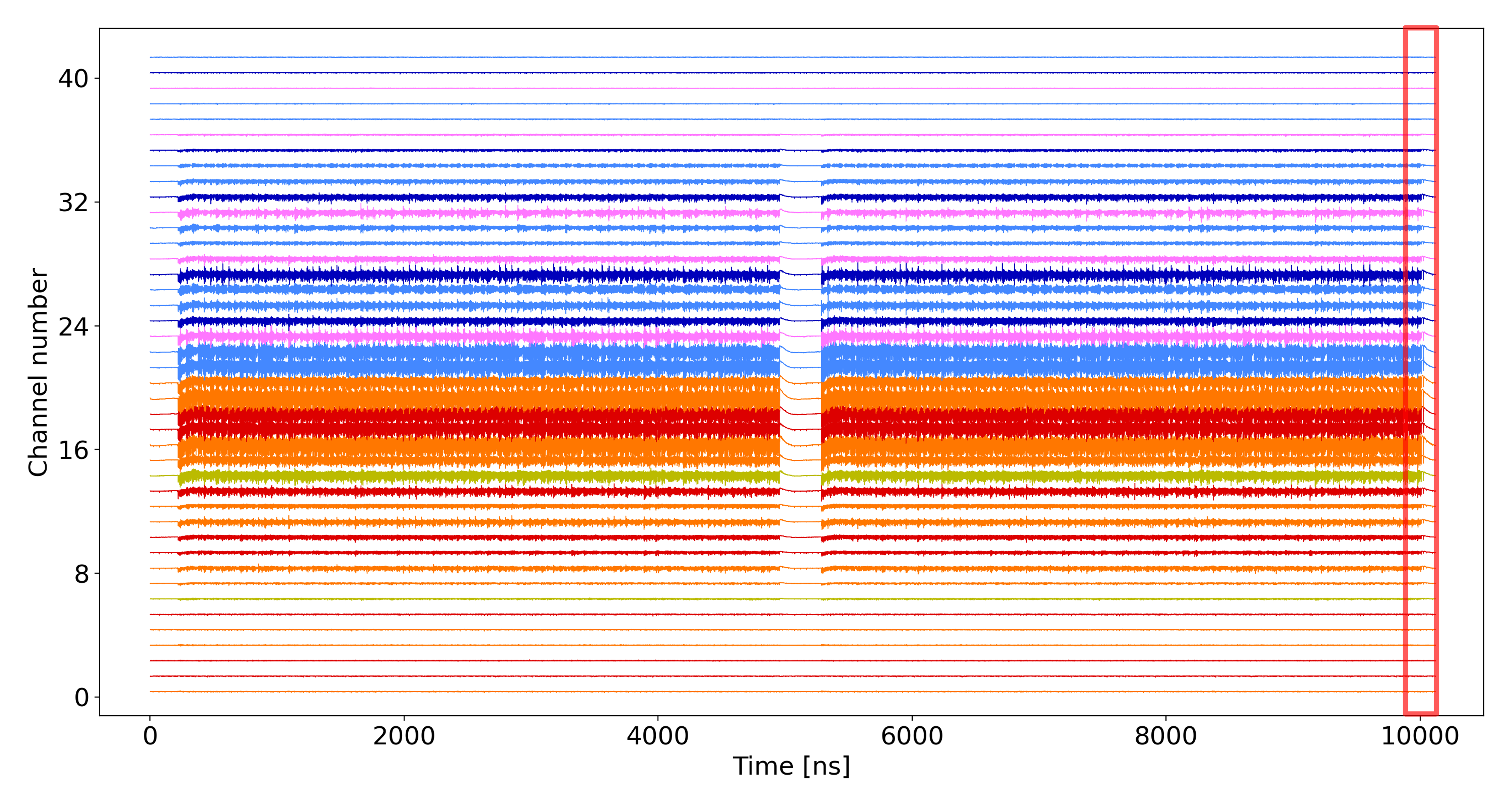}\\
    (a)
  \end{minipage}
  \vspace{0.8em}

  \begin{minipage}[b]{0.52\linewidth}
    \centering
    \includegraphics[width=\linewidth]{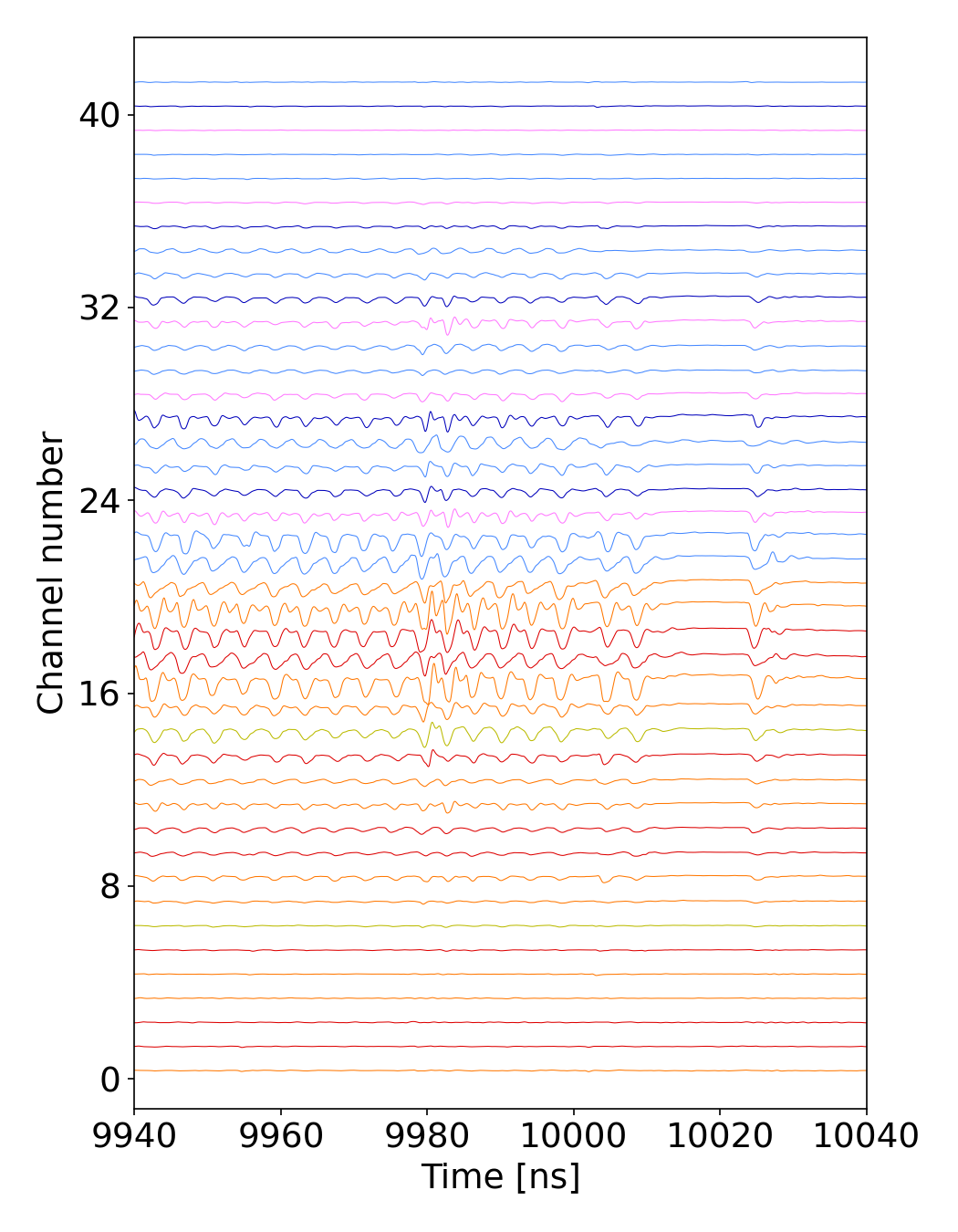}\\
    (b)
  \end{minipage}
  \caption{Raw SiXRM waveforms from an HER measurement acquired on February 2, 2026, with a beam current of 800~mA and 2346 bunches: (a) one full beam revolution; (b) magnified view of the red boxed region in (a). Waveforms from sensor channels 0--41 are stacked vertically as a function of time, and traces with the same color correspond to channels read out by the same ASIC.}
  \label{fig:2026-02-02.090248_waveform_overview}
\end{figure}

Figure~\ref{fig:2026-02-02.090248_waveform_overview} shows raw waveforms acquired from all sensor channels during a representative HER measurement.
The displayed interval corresponds to one full beam revolution, approximately $10\,\mu\mathrm{s}$.
In sweep-mode operation, waveforms are acquired repeatedly over short time windows corresponding to 12 RF buckets, and a full-turn waveform is reconstructed by concatenating the acquired segments.
Each window requires approximately 50~ms to acquire, giving a total measurement time of about 20~s per sweep.

Because signals are observed only at the times of filled RF buckets, the measured waveform directly reflects the bunch pattern.
In the SuperKEKB main ring, a typical fill pattern consists of two bunch trains separated by two gaps of approximately 300~ns.
These gaps, referred to as abort gaps, provide the time required for the beam-abort kicker to rise~\cite{Mimashi:IPAC2014-MOPRO023}.
The two bunch trains and the two abort gaps are visible in Fig.~\ref{fig:2026-02-02.090248_waveform_overview}(a).
The expanded view in Fig.~\ref{fig:2026-02-02.090248_waveform_overview}(b) resolves individual bunch signals as downward pulses.
The isolated pulse at the far right corresponds to the pilot bunch, which is separated from the main bunch train and used for betatron tune measurements without collisions.
The observation of the expected abort gaps and pilot bunch confirms that the measured waveform reproduces the beam timing structure.
By extracting the pulse height for each bunch in each channel and arranging the results spatially, the X-ray image formed through the coded aperture can be reconstructed.

\subsection{Peak Height Extraction}
\label{sec:Peak_Height_Extraction}
\begin{figure}[h]
  \centering
  \includegraphics[width=0.65\linewidth]{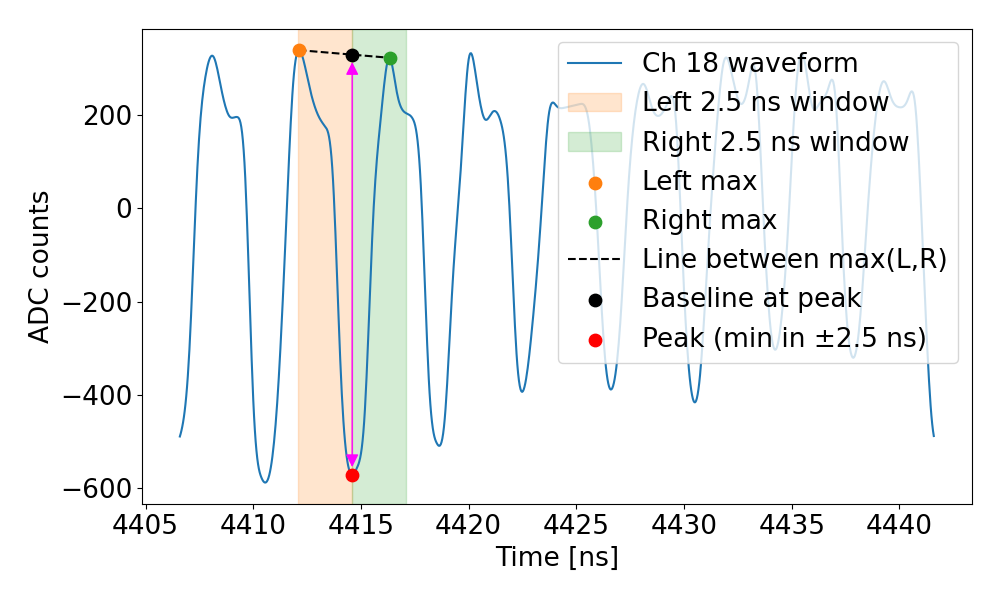}
  \caption{Illustration of the baseline and pulse-height determination for a single channel-18 pulse. The red point marks the pulse peak, the orange and green points indicate the maxima used to estimate the local baseline, and the magenta arrow shows the extracted pulse height.}
  \label{fig:2026-02-02.090248_baseline_height}
\end{figure}

As shown in Fig.~\ref{fig:2026-02-02.090248_waveform_overview}(b), the signals are often followed by overshoot and oscillatory ringing.
These waveform distortions can shift the local baseline of subsequent pulses, making a simple peak-to-baseline estimate unreliable.
The pulse height is therefore extracted using the following procedure:

\begin{enumerate}
\renewcommand{\labelenumi}{(\roman{enumi})}
\item \textbf{Reference-channel selection} \\
A reference channel is selected from channels with relatively large pulse amplitudes.
Channel 18 is typically used for this purpose.

\item \textbf{Spline interpolation} \\
Cubic spline interpolation is applied to all channel waveforms to obtain a smooth waveform representation.
The waveforms are upsampled by a factor of ten relative to the original sampling rate of 2.7~GSa/s.

\item \textbf{Fine timing alignment} \\
Small channel-to-channel timing offsets are corrected using the cross-correlation with the reference-channel waveform.
Each waveform is shifted within a range of $\pm 1\,\mathrm{ns}$ to maximize the correlation.

\item \textbf{Peak detection} \\
Pulse times are identified in the reference channel using the \texttt{find\_peaks} function from the SciPy library.
The same peak times are then used for all channels.

\item \textbf{Peak windowing} \\
For each detected pulse time, a $\pm 2.5\,\mathrm{ns}$ window is defined.
This window width is chosen with respect to the minimum bunch spacing in the SuperKEKB main ring, approximately 4~ns.
Within the window, the minimum value of each channel waveform is taken as the pulse peak.

\item \textbf{Baseline estimation} \\
The local baseline is estimated from the same $\pm 2.5\,\mathrm{ns}$ window.
The maximum values in the left and right halves of the window are connected by a straight line, and the value of this line at the pulse time is used as the baseline.
This procedure is illustrated in Fig.~\ref{fig:2026-02-02.090248_baseline_height}.

\item \textbf{Pulse-height calculation and calibration} \\
The pulse height is defined as the difference between the pulse peak and the estimated baseline.
The extracted amplitudes are then multiplied by the gain calibration coefficients obtained in Sec.~\ref{sec:Gain_Calibration}.
\end{enumerate}

\subsection{X-ray Image Reconstruction and Fitting}
\begin{figure}[h]
  \centering
  \includegraphics[width=0.70\linewidth]{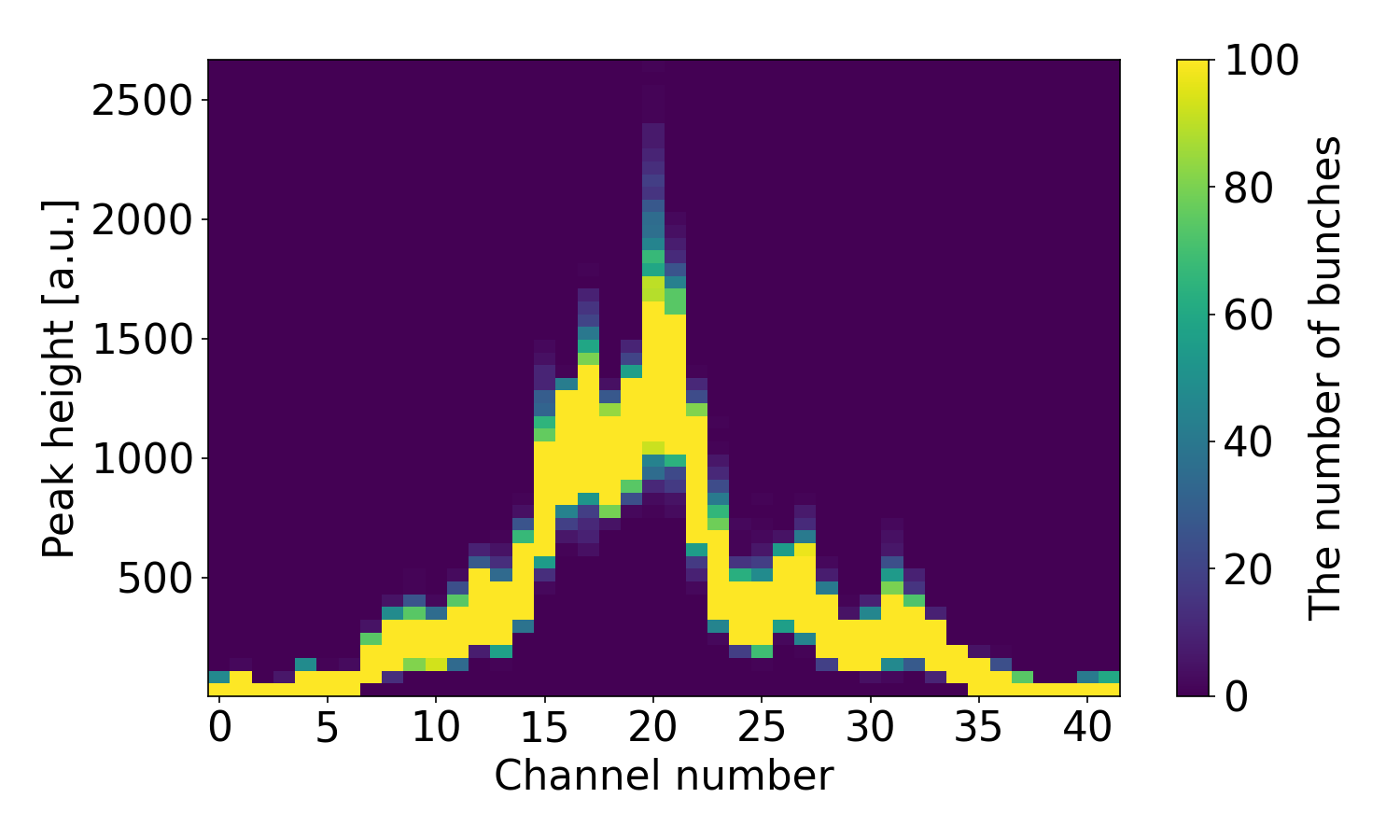}
  \caption{Superimposed reconstructed X-ray images for all bunches in one revolution. The horizontal axis is the silicon sensor channel number, the vertical axis is the calibrated pulse height, and the color scale indicates the number of bunches in each bin.}
  \label{fig:2026-02-02.090248_peakheight_heatmap}
\end{figure}

For each bunch, the calibrated pulse heights from the 42 sensor channels form an X-ray image along the vertical detector coordinate.
Figure~\ref{fig:2026-02-02.090248_peakheight_heatmap} shows the distribution obtained by superimposing the reconstructed images for all bunches in the representative HER measurement shown in Fig.~\ref{fig:2026-02-02.090248_waveform_overview}.
The coded-aperture peak structure is clearly visible.

The vertical beam size is then determined by fitting the reconstructed X-ray image for each bunch.
The same coded-aperture response templates and fitting method as those used for the existing CMOS-based XRM system~\cite{Mulyani:2019gsy} are adopted, enabling a direct comparison between the two systems.
The least-squares fit includes six free parameters,
$\theta=(\mu, \sigma, v_{\mathrm{offset}}, x_0, \mathrm{norm}, \mathrm{scale})$,
where $\mu$ and $\sigma$ denote the vertical beam position and vertical beam size, $v_{\mathrm{offset}}$ and $x_0$ describe image-position offsets, and $\mathrm{norm}$ and $\mathrm{scale}$ account for image-amplitude and horizontal-scale variations, respectively.
In the present analysis, $\sigma$ is the parameter of interest, while the remaining parameters are treated as nuisance parameters that absorb position, alignment, and normalization variations.
For a given parameter set $\theta$, the template model predicts the pulse height $y^{\mathrm{model}}_i(\theta)$ at sensor channel $i$.

The fit quality is evaluated using the sum of squared residuals, denoted here as $\chi^2$ and defined as
\begin{equation}
\chi^2(\theta) = \sum_{i=0}^{41} \left( y^{\mathrm{model}}_i(\theta) - y^{\mathrm{data}}_i \right)^2,
\label{eq:chi^2}
\end{equation}
where $y^{\mathrm{data}}_i$ denotes the measured pulse height at channel $i$.

\begin{figure}[h]
  \centering
  \includegraphics[width=0.68\linewidth]{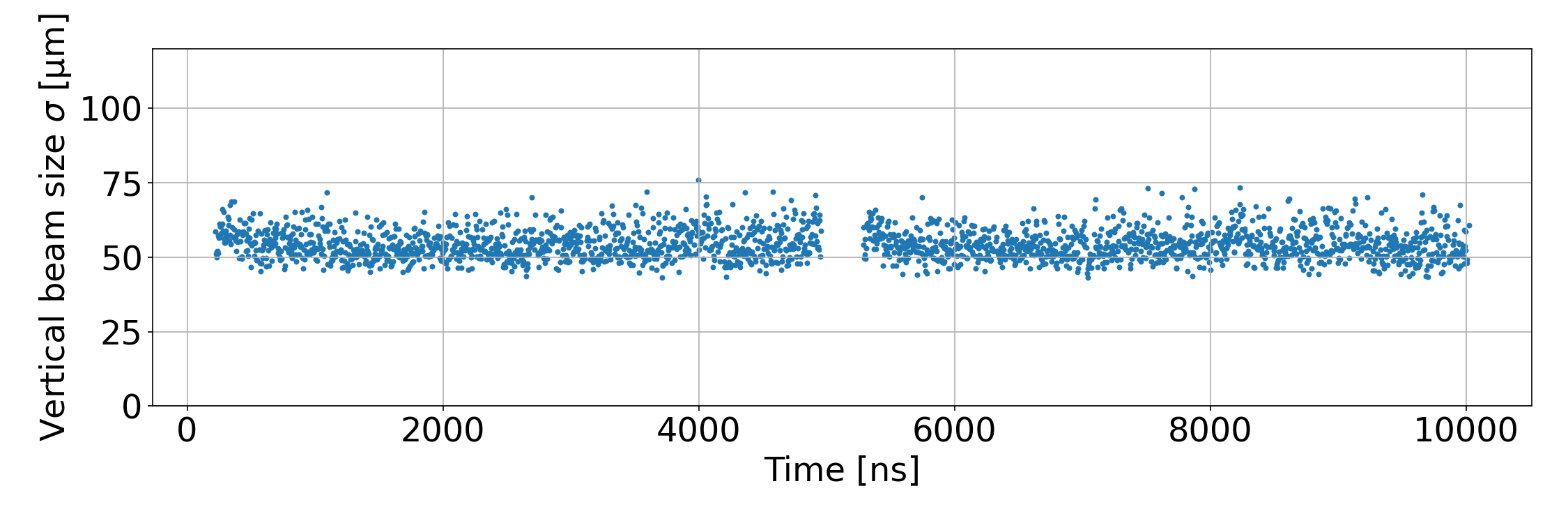}
  \caption{Bunch-by-bunch vertical beam size obtained from template fitting for one full revolution. Each point corresponds to one bunch; the gaps correspond to the abort gaps in the fill pattern.}
  \label{fig:2026-02-02.090248_fitresult_sigma}
\end{figure}

\begin{figure}[h]
  \centering
  \includegraphics[width=0.65\linewidth]{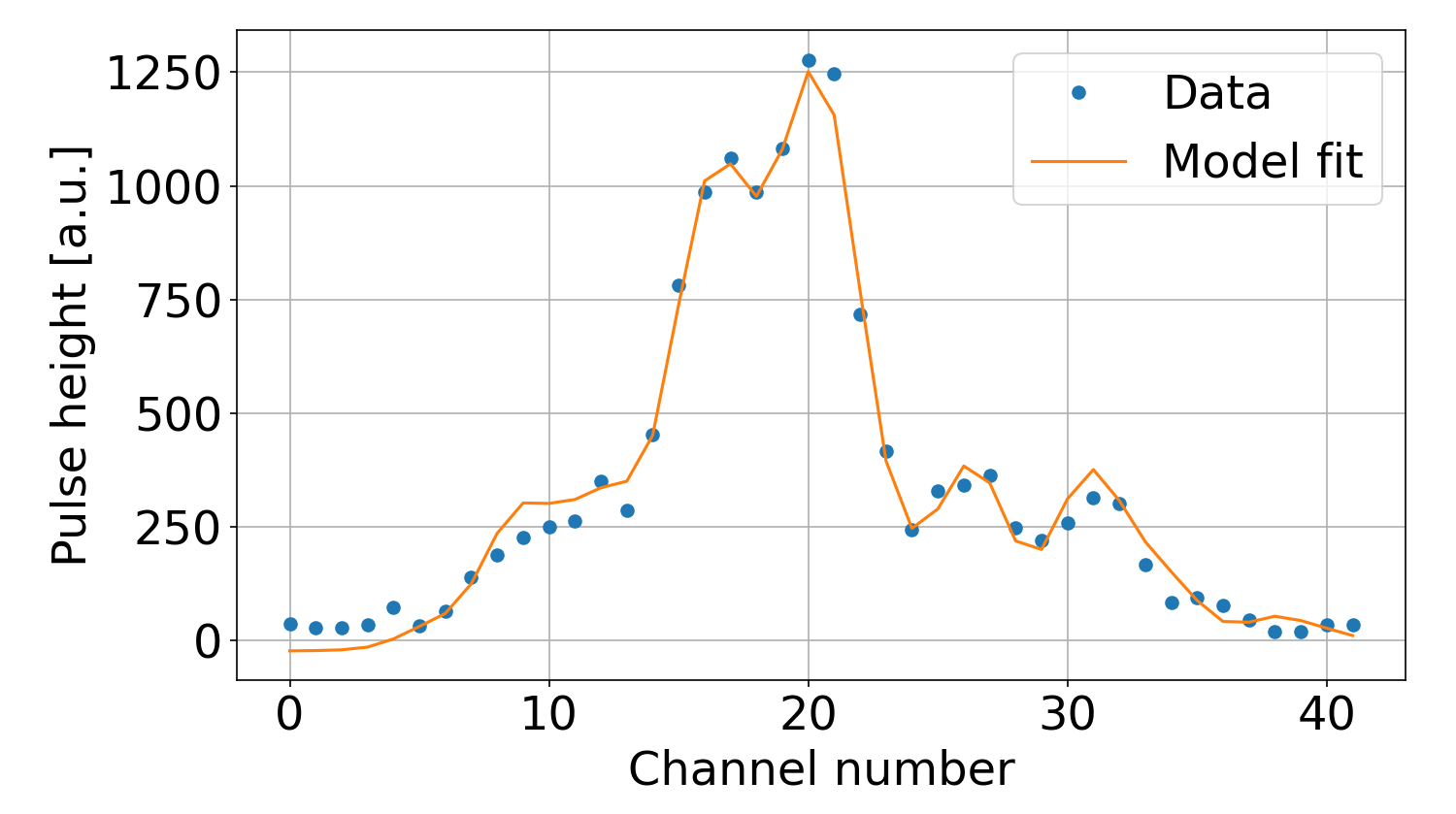}
  \caption{Example template fit to a single-bunch X-ray image. The blue points show the measured pulse heights, and the orange curve shows the fitted template.}
  \label{fig:2026-02-02.090248_examplefit}
\end{figure}

\begin{figure}[h]
  \centering
  \includegraphics[width=0.68\linewidth]{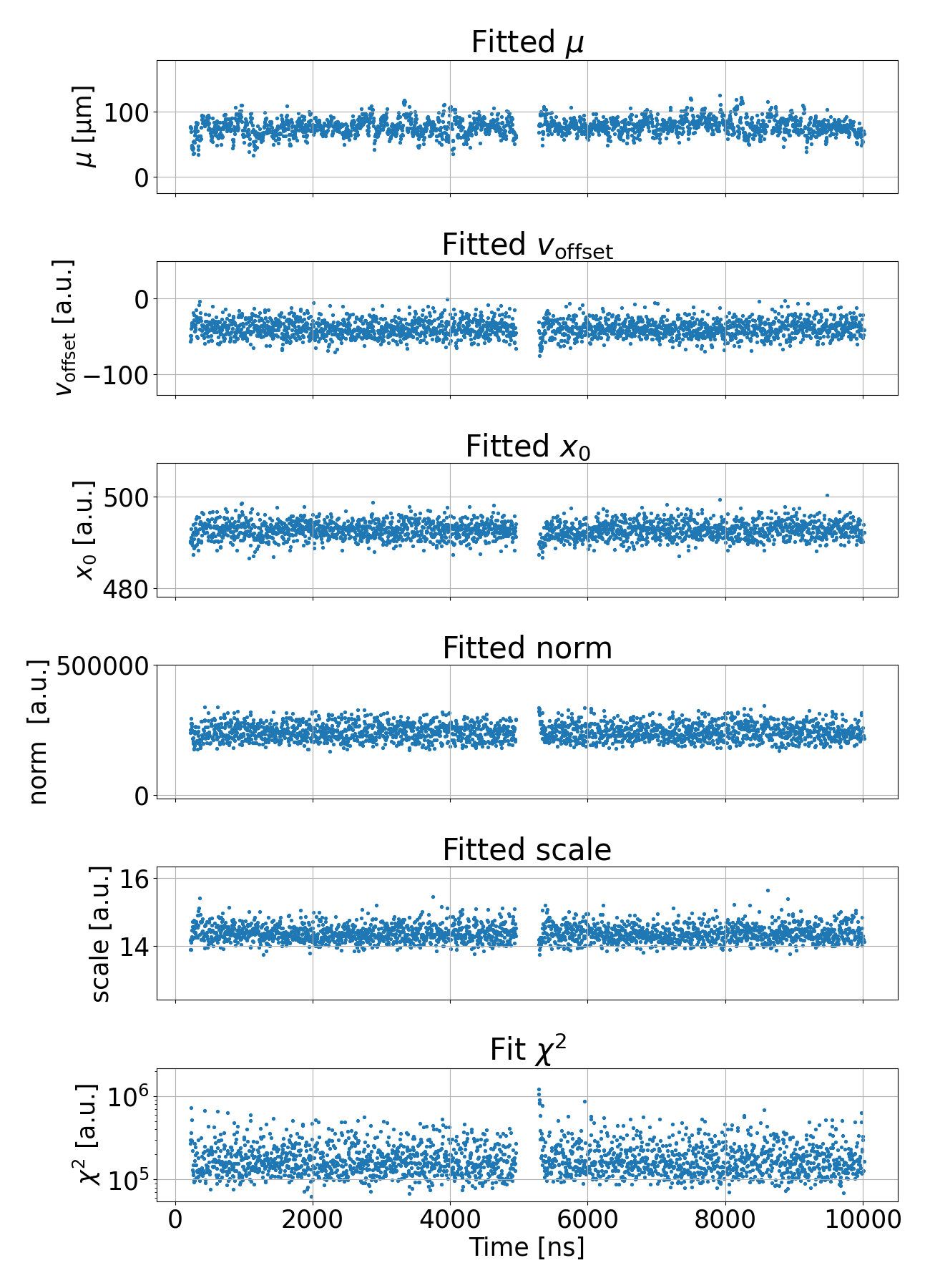}
  \caption{Fitted nuisance parameters and corresponding $\chi^2$ values for the bunch-by-bunch template fits. All panels share the same horizontal axis as Fig.~\ref{fig:2026-02-02.090248_fitresult_sigma}.}
  \label{fig:2026-02-02.090248_fitresult_summary}
\end{figure}

Figure~\ref{fig:2026-02-02.090248_fitresult_sigma} shows the bunch-by-bunch fitted vertical beam sizes for the representative measurement.
The average over all bunches in one turn is 54.3~$\mu\mathrm{m}$, while the corresponding value measured by the CMOS-based XRM is 56.6~$\mu\mathrm{m}$.
The agreement is reasonable, given that the SiXRM value is obtained bunch by bunch whereas the CMOS-based XRM provides an exposure-averaged measurement.
An example fit to a single-bunch X-ray image is shown in Fig.~\ref{fig:2026-02-02.090248_examplefit}.
The remaining fitted parameters and the corresponding $\chi^2$ values are summarized in Fig.~\ref{fig:2026-02-02.090248_fitresult_summary}.
The fitted parameters are stable over the revolution, although the first few bunches at the head of the bunch train show larger $\chi^2$ values, attributed to baseline fluctuations in the waveform.

\section{Performance Evaluation} 

The SiXRM measurement performance is evaluated using HER beam data and compared with the existing CMOS-based XRM system. The comparison validates the reconstructed beam sizes and provides an estimate of the measurement precision.

\subsection{Validation with the existing CMOS-based XRM system}
The existing CMOS-based XRM system~\cite{Mulyani:2019gsy} operates continuously during SuperKEKB machine operation and provides reference measurements of the vertical beam size.
A comparison with this system provides a direct validation of the SiXRM reconstruction.
A total of 2429 SiXRM sweep measurements were acquired from late 2025 to early 2026 under various accelerator conditions, covering a wide range of vertical beam sizes.

\begin{figure}[h]
  \centering
  \includegraphics[width=0.62\linewidth]{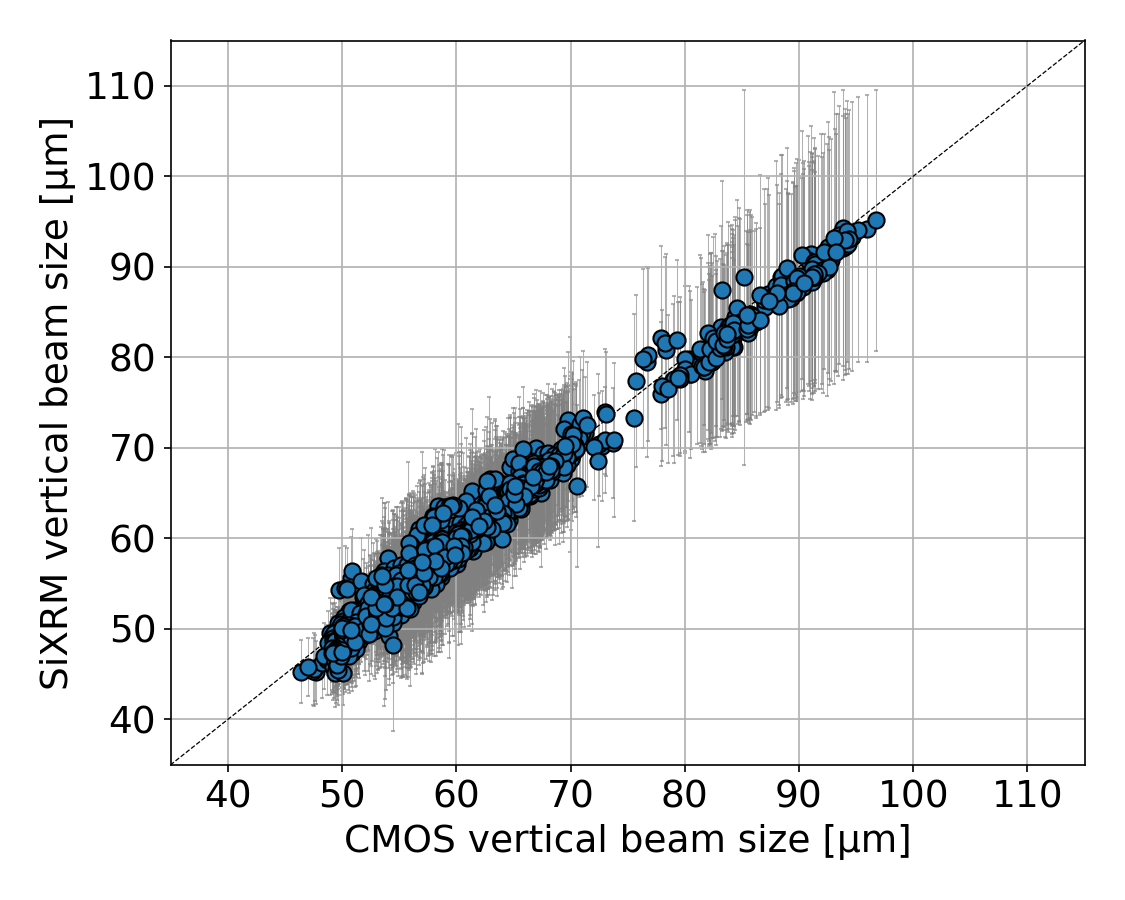}
  \caption{Comparison of the vertical beam size measured by the SiXRM and the existing CMOS-based XRM for 2429 sweep measurements. Each point represents one SiXRM sweep measurement.}
  \label{fig:CMOS_vs_SiXRM}
\end{figure}

Figure~\ref{fig:CMOS_vs_SiXRM} compares the bunch-averaged vertical beam size measured by the SiXRM with the corresponding value measured by the CMOS-based XRM.
Each point corresponds to one SiXRM sweep measurement.
The SiXRM value is defined as the average of the bunch-by-bunch fitted beam sizes over one revolution, and the error bar indicates the standard deviation over bunches in that revolution.
The CMOS-based XRM value is defined as the average beam size over a 20~s interval starting from the beginning of the corresponding SiXRM sweep.
Since the CMOS-based XRM records beam-size measurements at intervals of approximately 1~s, this average is obtained from about 20 CMOS measurements.
The measurement resolution of the CMOS-based XRM is approximately 1~$\mu\mathrm{m}$, expressed as a one-standard-deviation value.
The data points lie close to the diagonal line, indicating good agreement between the two systems and validating the SiXRM beam-size reconstruction.

\subsection{Estimate of the SiXRM measurement precision}
Because the CMOS-based XRM has a measurement precision of approximately 1~$\mu\mathrm{m}$, it provides a suitable reference for evaluating the SiXRM measurement precision.
For each SiXRM sweep measurement $i$, the vertical beam size of bunch $j$, $\sigma^{\mathrm{SiXRM}}_{ij}$, is compared with the corresponding CMOS-based XRM value $\sigma^{\mathrm{CMOS}}_{i}$.
Here, $\sigma^{\mathrm{CMOS}}_{i}$ is defined as the average CMOS-based XRM beam size over the 20~s time window corresponding to the SiXRM sweep.

The residual for each bunch is defined as
\begin{equation}
r_{ij}=\sigma^{\mathrm{SiXRM}}_{ij}-\sigma^{\mathrm{CMOS}}_{i}.
\end{equation}
Thus, one residual is obtained for each bunch in each sweep measurement.
The SiXRM precision estimator is then defined as the standard deviation of the residual distribution over all bunches and all sweep measurements:
\begin{equation}
\label{eq:precision}
\sigma_{\mathrm{prec}}=\sqrt{\frac{1}{\sum_{i=1}^{N_{\mathrm{evt}}} N_i}\sum_{i=1}^{N_{\mathrm{evt}}}\sum_{j=1}^{N_i}\left(r_{ij}-\bar{r}\right)^2},
\end{equation}
where $N_{\mathrm{evt}}$ is the total number of SiXRM sweep measurements, and $N_i$ is the number of detected bunches in the $i$-th sweep measurement.
The total number of data points is therefore given by $\sum_{i=1}^{N_{\mathrm{evt}}} N_i$.
Here, $\bar{r}$ denotes the mean value of the residuals over all bunches and all sweep measurements.

Applying this method to the 2429 sweep measurements gives $\sigma_{\mathrm{prec}} = 6.4~\mu\mathrm{m}$.
This value includes the intrinsic SiXRM measurement resolution.
It also includes contributions from machine variations and bunch-by-bunch beam-size variations.
It therefore should not be interpreted as the intrinsic detector resolution alone.
Rather, it provides a conservative upper bound on the SiXRM measurement uncertainty.
The vertical beam-size measurement precision of the SiXRM system is thus estimated to be better than 6.4~$\mu$m.

\FloatBarrier

\section{Future Work} 

The performance evaluation demonstrates that the present SiXRM system can provide reliable bunch-resolved vertical beam-size measurements using synchrotron X-rays at SuperKEKB. It also identifies several areas where further development can improve the measurement capability. The next development steps are aimed at increasing the readout speed, reducing waveform-related systematic uncertainties, and extending the technique to broader operating conditions.

A primary limitation of the current implementation is the sweep-based readout scheme, which reconstructs a full revolution from multiple short acquisitions rather than recording the full turn simultaneously. Future upgrades will focus on readout architectures capable of true full-turn, bunch-by-bunch acquisition. One promising direction is the use of high-speed digitization platforms such as AMD/Xilinx RFSoC devices, which integrate fast analog-to-digital converters and FPGA logic in a single device. Improvements to the existing boardstack, especially in the digitization stage, may also reduce conversion time and expand the accessible time window.

Further improvements are also planned for online processing. Implementing real-time feature extraction in the readout system would reduce the data volume by avoiding full waveform storage for every acquisition. This would simplify downstream analysis and improve operational usability during machine studies.

Finally, extension of the SiXRM technique to the LER is an important next step. Because the lower beam energy is expected to result in smaller X-ray signals, this application will require improved front-end electronics with higher gain and lower noise. Alternative sensor technologies, including InGaAs in addition to silicon, will also be explored to improve sensitivity.

\section{Conclusion} 

We have developed and commissioned a silicon-based X-ray radiation monitor for bunch-resolved vertical beam-size measurements at SuperKEKB. The system combines a silicon strip sensor, amplifier boards, and boardstack-based waveform readout electronics to reconstruct X-ray images for individual bunches.

Beam measurements in the HER demonstrate clear reconstruction of the bunch structure and coded-aperture X-ray images. The bunch-averaged beam sizes obtained with the SiXRM agree well with those measured by the existing CMOS-based XRM system, validating the reconstruction method. Using the CMOS-based XRM as a reference, the SiXRM measurement precision is estimated to be better than 6.4~$\mu$m as a conservative upper bound.

These results establish the feasibility of X-ray-based bunch-resolved beam-size measurements at SuperKEKB. The SiXRM provides a new diagnostic capability for studying beam-size evolution along the bunch train and supporting machine studies toward higher luminosity.

\section{Acknowledgements} 
This work was supported in part by the U.S. Department of Energy (DOE), Office of High Energy Physics, under Grant No.~DE-SC0010504.
Additional support was provided by the U.S.--Japan Science and Technology Cooperation Program in High Energy Physics.
It was also supported by JSPS International Leading Research under Grant No.~JP22K21347.

The authors thank the SuperKEKB accelerator group for stable beam operation and continuous support during the beam studies.
We also thank the Belle~II collaboration for valuable discussions and technical support.

\bibliographystyle{elsarticle-num}
\bibliography{references}

\end{document}